\documentclass[a4paper,11pt]{article}
\usepackage{url}
\usepackage{jinstpub} 
\usepackage{lineno}

\title{\boldmath First characterization of a novel grain calorimeter: the GRAiNITA prototype}

\author[a]{Sergey Barsuk,}
\author[d]{Oleg Bezshyyko,}
\author[a,b]{Ianina Boyarintseva,}
\author[b]{Andrey Boyarintsev,}
\author[a]{Dominique Breton,}
\author[c]{Herv\'e Chanal,}
\author[b]{Alexander M. Dubovik,}
\author[d]{Andrii Kotenko,}
\author[a]{Giulia Hull,}
\author[a]{Jacques Lefran\c{c}ois,}
\author[c]{St\'ephane Monteil,}
\author[a,1]{Marie-H\'el\`ene Schune,\note{Corresponding author.}}
\author[d]{Nazar Semkiv,}
\author[b]{Irina Tupitsyna}
\author[c]{and Mykhailo Yeresko.}
\affiliation[a]{Université Paris-Saclay, CNRS/IN2P3, IJCLab, Orsay, France}
\affiliation[b]{Institute for Scintillation Materials of the National Academy of Sciences of Ukraine, 60 Nauki Ave., Kharkiv 61072, Ukraine}
\affiliation[c]{Université Clermont-Auvergne, CNRS/IN2P3, LP-Clermont, 63177 Aubiere, France}
\affiliation[d]{Kyiv National Taras Shevchenko University, 01033 Kyiv, Ukraine}

\emailAdd{marie-helene.schune@ijclab.in2p3.fr}

\abstract{A  novel type of calorimeter based on grains of inorganic scintillating crystal readout by wave length shifting fibers is proposed. The concept and main features as well as the prototype design are introduced and the first results obtained using cosmic rays are presented. The number of photo-electrons generated by cosmic rays muons in the prototype detector is estimated to be of the order of 10000 photo-electrons per GeV, validating the concept of this next-generation shashlik calorimeter.}


\arxivnumber{2312.07365} 

\begin{document}
\maketitle
\flushbottom

\section{Introduction}
Electromagnetic calorimeters in High Energy Physics face a dilemma as the use of uniform inorganic crystals, while providing excellent energy resolution, comes at a high cost and with limited granularity and lateral non-uniformity. On the other hand, more affordable sampling calorimeters have limited energy resolution due to their inherent limitations such as the presence of passive absorbers limiting the sampling fraction.

In a shashlik-type sampling calorimeter a stack of alternating slices of absorber and scintillator materials is penetrated by wavelength shifting fibers (WLS) running perpendicular to the absorber and scintillator tiles, for the light collection towards photodetectors. Within the GRAiNITA concept, it is proposed to mix in the same volume high-Z scintillator grains and a high-density transparent liquid as absorber. The multiple refractions of the light on the grains ensures the stochastic confinement of the light, as in the LiquidO detection technique~\cite{LiquidO}. The scintillation light can be collected towards the photodetectors by means of WLS fibers, regularly distributed in the detection volume as for a  conventional shashlik detector.
Due to this extremely fine sampling of the electromagnetic shower, an excellent energy resolution is expected.

\subsection{Detector concept}
\begin{figure}[htbp]
\centering
\includegraphics[width=.6\textwidth]{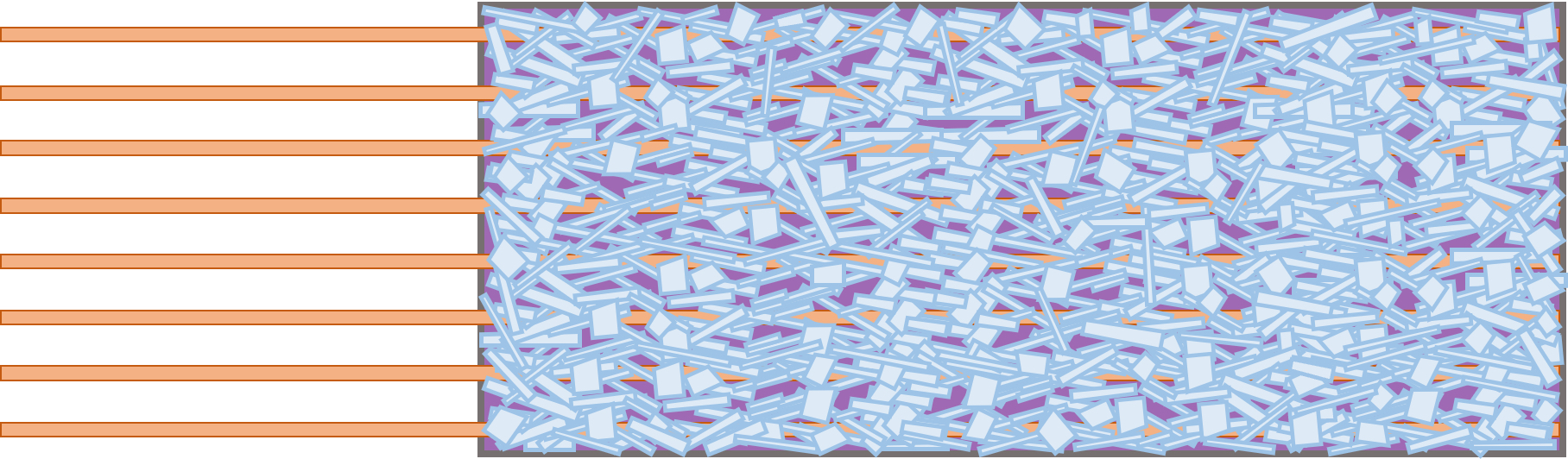}
\hskip .2cm
\includegraphics[width=.2\textwidth]{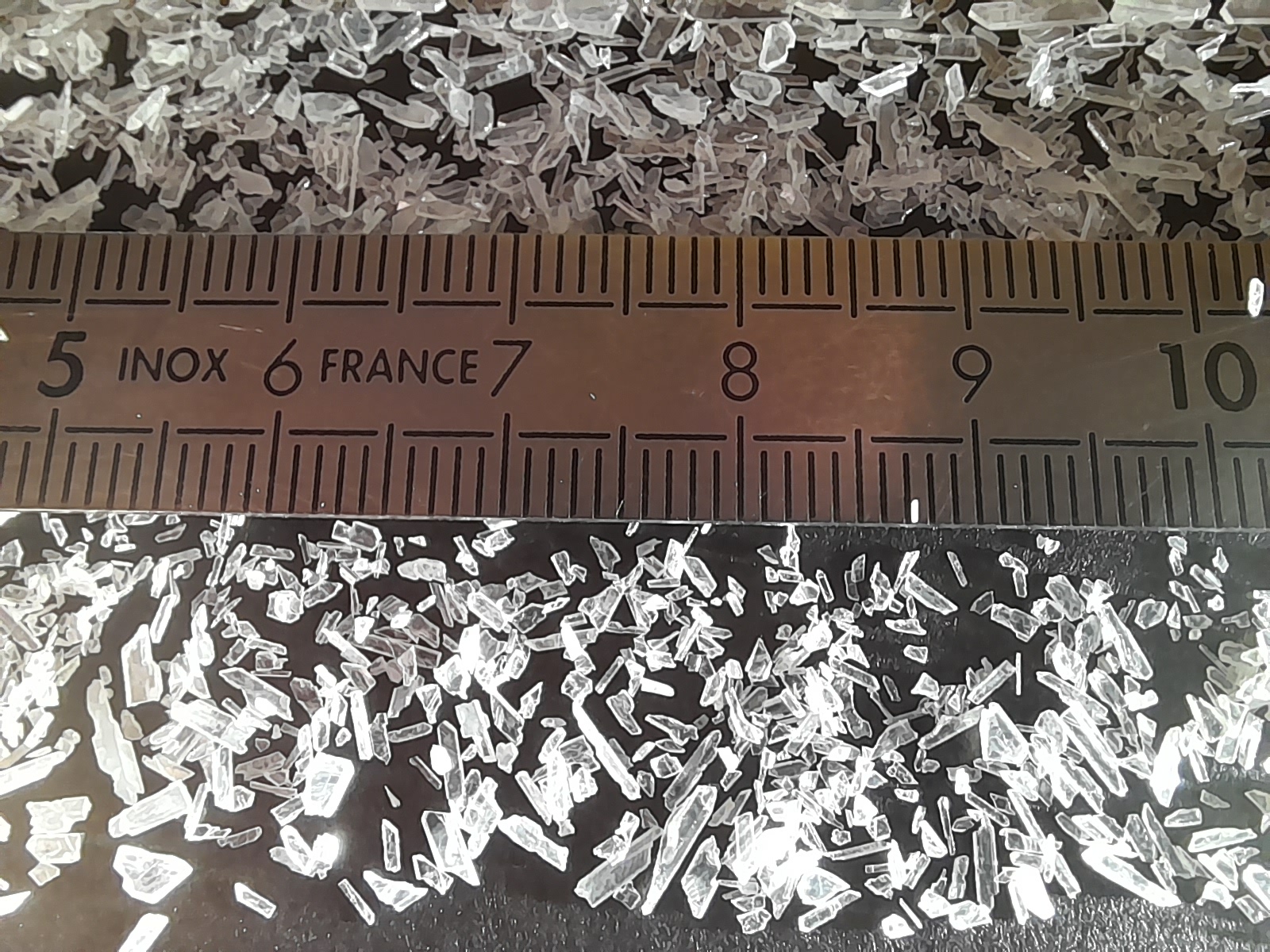}
\caption{Left: sketch of the GRAiNITA calorimeter. The (light blue) scintillating grains are immersed in a (purple) liquid and the WLS fibers are represented in orange. Right: picture of the ZnWO$_4$ crystal grains produced at ISMA.\label{fig:concept}}
\end{figure}

The basic idea of this new type of calorimeter is to use high-Z scintillator grains with typical size of a fraction of millimeter. The scintillation light can then be collected towards the photodetectors, in particular SiPMs, by means of WLS fibers, as in a conventional shashlik-type calorimeter. An addition of high-density and high refractive-index liquid in the detection volume not only increases the overall density but assures better optical matching with the scintillating grains, thus controlling the refraction and reflection interactions of the light, in the diffusion towards the WLS fibers. A schematic view of the detector concept is shown in figure~\ref{fig:concept}-left. Two high-density and high-Z scintillator materials, ZnWO$_4$ and BGO \-- both excellent candidates for our studies \--  are used for the development of the GRAiNITA prototypes described in this article.

\subsection{Preliminary simulation\label{sec:Geant4MC}}
First Geant4 Monte-Carlo~\cite{G4} studies
were performed for a 1 GeV photon sent to a simplified GRAiNITA calorimeter consisting of the mix of 2 mm side cubes of scintillators (ZnWO$_4$) and absorbers (for this study CH$_2$ I$_2)$, randomly distributed.
In this study, for comparison, a typical shaslik calorimeter with a geometry similar to the LHCb experiment one was also simulated. It consists in alternating layers of 20 mm $\times$ 20 mm $\times$ 2 mm lead and scintillating plates.
In both simulated cases, the overall thickness of the calorimeters was chosen in such a way that there are at least 25 radiation lengths in depth of the absorber material.
With this simple simulation, the energy resolution obtained for the LHCb-like calorimeter is
$\frac{\sigma_E}{E} = \sim \frac{9\%}{\sqrt{E}}$,  in agreement with the LHCb results~\cite{LHCbSergey,LHCbCalo}.  The result for the randomly-distributed 2-mm cubes is $\frac{\sigma_E}{E} = \sim \frac{3\%}{\sqrt{E}}$. Since the energy resolution is expected to decrease with the grain size, a stochastic energy resolution of the order of $\frac{\sigma_E}{E} = \frac{2\%}{\sqrt{E}}$ could be at reach.
If it is indeed the case, the GRAiNITA calorimeter could be used in the context of a future $e^+ e^-$ collider. It would be particularly interesting for Flavour Physics experiments running at the $Z^0$ pole as at FCC-ee~\cite{bib:FCCee} for rare decays involving photons or neutral pions in the final state. Since the Z$^0$ production rate at FCC-ee is of the order of 10$^5$ Hz and since these events should typically hit less than 1 \% of the calorimeter cells, the  20 $\mu$s decay time of the ZnWO$_4$ should not be a concern. More details on the simulation and feasibility study of GRAiNITA are available in ~\cite{bib:IEEE2021}.

\section{Choice of detector components and light propagation studies}
\subsection{Scintillating grains}
ZnWO$_4$ and BGO are well known high-density and high-Z scintillator materials, and both are excellent candidates for the GRAiNITA calorimeter.
Their properties of interest for this study are summarized in table~\ref{tab:properties}.
\begin{table}[htbp]
\centering
\caption{Properties of interest for the GRAiNITA study of the the BGO and ZnWO$_4$ materials.\label{tab:properties}}
\smallskip
\begin{tabular}{l|l|l}
\hline
    & BGO   & ZnWO$_4$ \\
  \hline
  Effective $Z$     &74 & 61\\
Density     $(g/cm^3)$  & 7.13       & 7.87 \\
Refractive index    & 2.15   & 2.0 \-- 2.3 \\
Light yield  (photons/MeV)  &  $\sim$ 9000     &$\sim$ 9000\\
Peak emission wavelength  (nm)   & 480 & 480 \\
Decay time  ($\mu s$)               & 0.3 & 20 \\
  Radiation length    (cm)       &  1.12  &   1.20 \\
  Moli\`ere radius     (cm)         & 2.26 & 1.98 \\
\hline
\end{tabular}
\end{table}
Since ZnWO$_4$ grains of the desired size can be successfully grown with the method of spontaneous crystallization from a flux melt~\cite{bib:Yamada} a special emphasis was put on this material. However for reference purposes, BGO grains, obtained crushing a uniform crystal into grains, via the planetary ball milling technique, were also studied.
These two types of scintillating grains have been produced at the Institute for Scintillation Materials (ISMA) in Kharkiv (Ukraine). While the production process can still be optimized, reasonable quantity (1 kg) of ZnWO$_4$ grains has already been produced and their quality satisfies the requirements of this study~\cite{bib:IEEE2022}. A picture of the grains is shown in figure~\ref{fig:concept}-right.

\subsection{Wave length shifting fiber selection }
\begin{figure}[htbp]
\centering
\includegraphics[angle=270,width=1.15\textwidth]{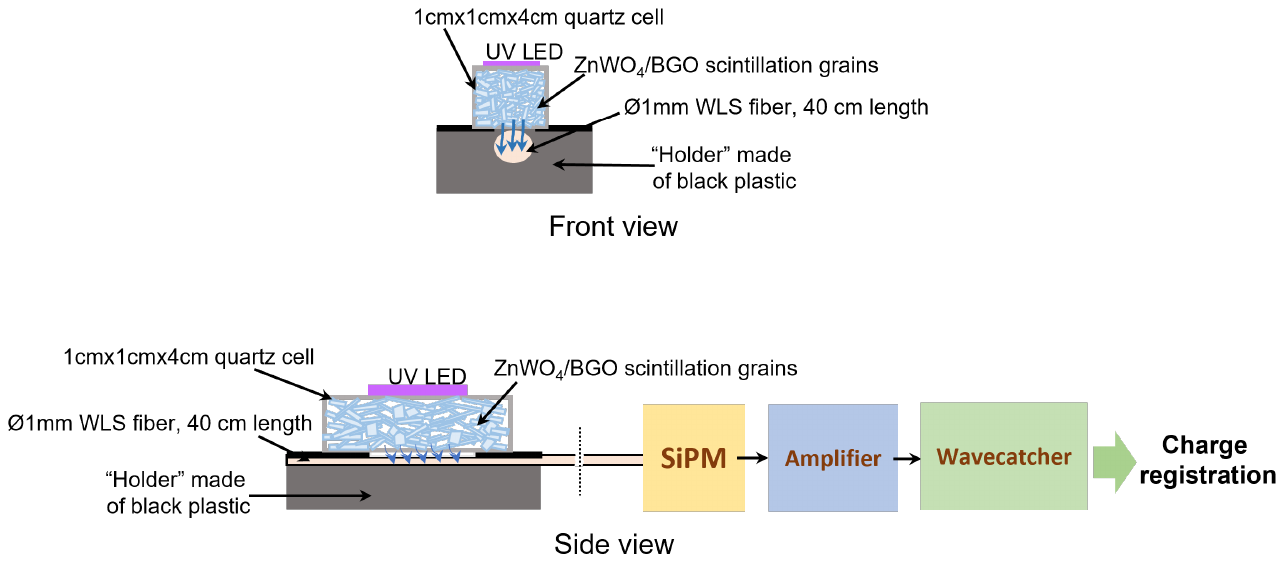}
\vskip -3cm
\caption{Sketch of the test bench used for testing the matching of different WLS fibers and grains.\label{fig:testWLS}}
\end{figure}

\begin{figure}[htbp]
\centering
\includegraphics[angle=270,width=.80\textwidth]{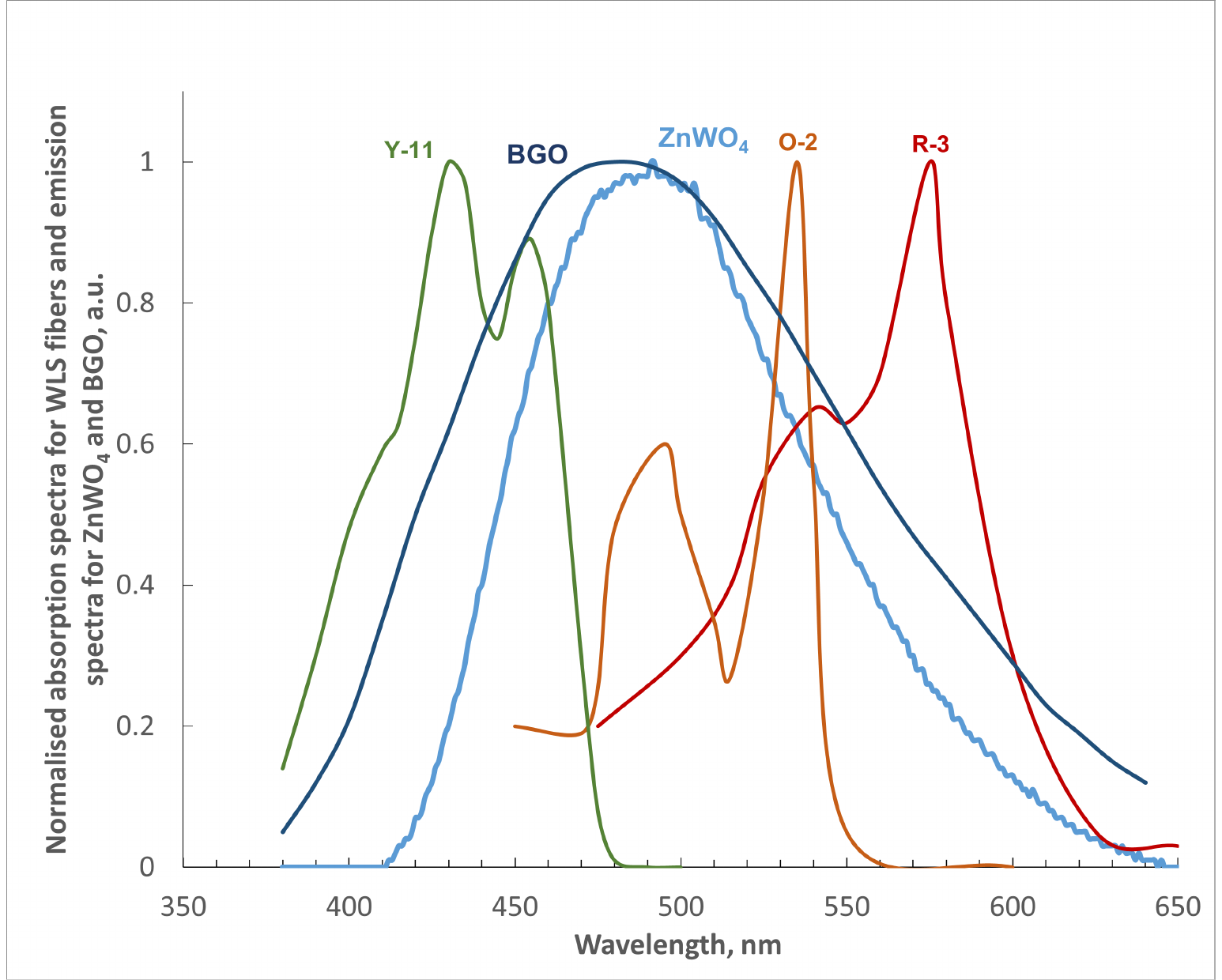}
\vskip -.25cm
\caption{Absorption spectra for Y-11, O-2 and R-3 fibers from Kuraray~\cite{bib:KurarayWLSOnline} and emission spectra for BGO~\cite{bib:SaintGobainBGOOnline} and ZnWO$_4$ measured at ISMA for this study.\label{fig:WLSspectra}}
\end{figure}

Using a dedicated setup with ZnWO$_4$ or BGO scintillators (see figure~\ref{fig:testWLS} for a sketch of the test bench), four WLS fibers from Kuraray, O-2(300), O-2(200), Y11(200) and R-3(100), with different  absorption spectra ~\cite{bib:KurarayWLSOnline} or doping level, are tested to find the best match to the emission spectra of the grains. For this test a small quartz cell is filled with scintillator grains (the thickness of the grains layer is about 9 mm) and illuminated with a UV LED\footnote{VISHAY VLMU35CL20-275-120 from Farnell}. The grains are thus excited and produce luminescence; the light is then collected with different WLS fibers towards a SiPM\footnote{  \href{https://www.hamamatsu.com/content/dam/hamamatsu-photonics/sites/documents/99_SALES_LIBRARY/ssd/s13360_series_kapd1052e.pdf}{S13360-1350pe from Hamamatsu}}. In this configuration, the produced light has a different intensity than that produced by ionizing radiation but very similar wavelength of emission. The absorption spectra for various WLS fibers are compared with X-ray excited emission spectra for BGO~\cite{bib:SaintGobainBGOOnline} and ZnWO$_4$ (measured at ISMA for this study) on figure~\ref{fig:WLSspectra}. The relative efficiency compared to the O-2(300) fiber is given in table~\ref{tab:fibers}. The best matching WLS is found to be the O-2(200), both for ZnWO$_4$ and BGO and is used for the cosmic test described below.

\begin{table}[htbp]
    \centering
    \begin{tabular}{l|l|l}
   & \multicolumn{2}{c}{Relative efficiency (\%)}  \\
  \hline
   Fiber type     &   ZnWO$_4$ grains     &  BGO  grains  \\
  \hline
  O-2(300) & 100 & 100 \\
  O-2(200) & 104 & 104 \\
  Y-11(200) & 44    &  98 \\
  R-3(100)  & 60    & n.a. \\
    \end{tabular}
    \caption{Relative efficiencies measured for several grains and WLS fibers pairing.}
    \label{tab:fibers}
\end{table}

\subsection{Light propagation studies}
\begin{figure}[htbp]
\centering
\includegraphics[width=.5\textwidth]{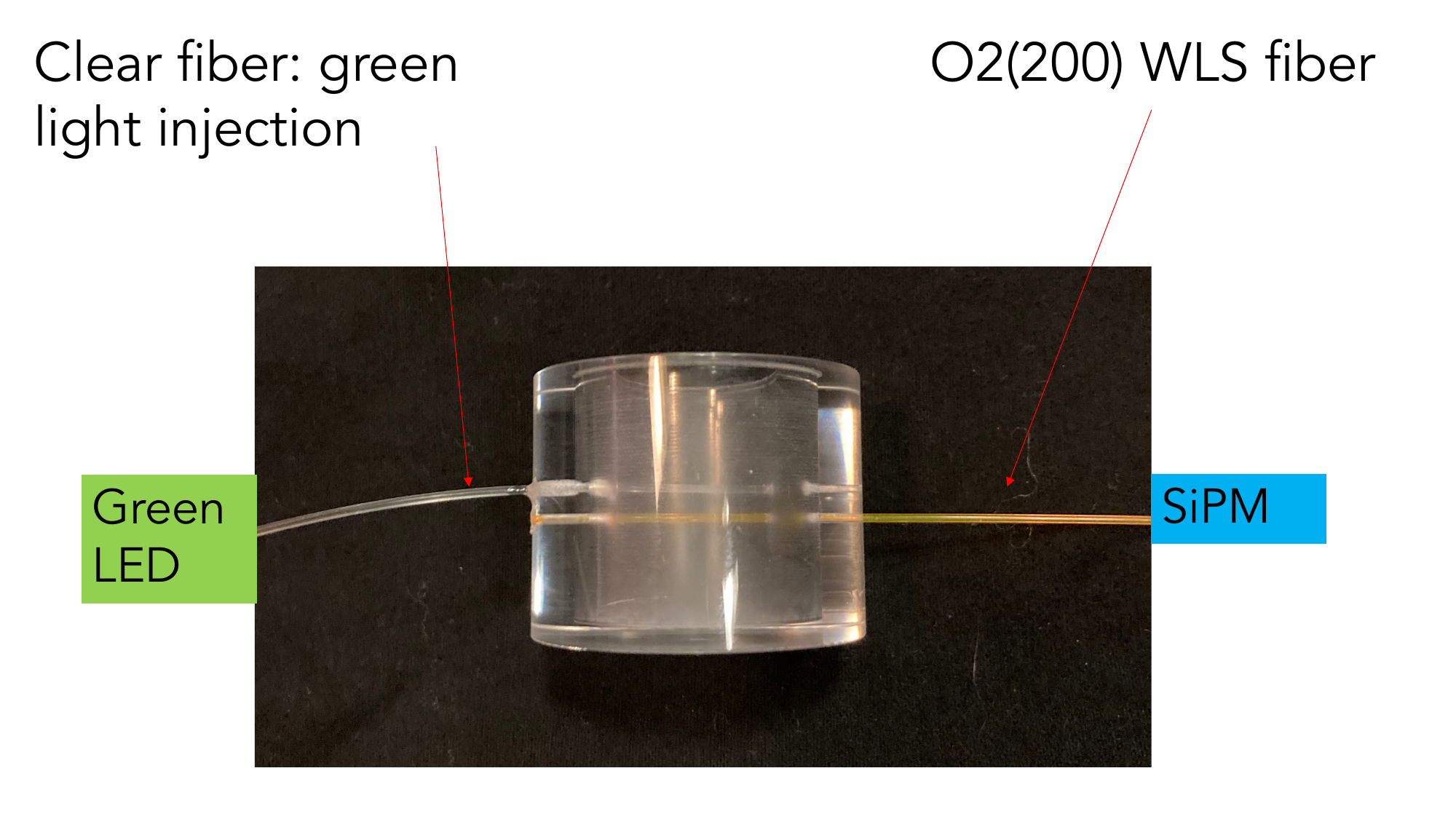}
\caption{Sketch of the light propagation test setup.\label{fig:LightPropFig}}
\end{figure}

In order to further evaluate the matching between the scintillating grains read-out by the O-2(200) fiber and the effect of the liquid on the light propagation we used a small GRAiNITA prototype. It consists in a small container, that can fit less than 50 g of grains, where two fibers, a WLS one and a clear one, are placed horizontally 4 mm away. The clear fiber is connected to a green light source and it is unpolished for 1 cm along its length to inject the light in the reference volume, while the O-2(200) fiber is coupled to a SiPM (see figure~\ref{fig:LightPropFig}). With this set-up we measured the charge collected at the photodetector when the container was, in turn, wrapped with black fabric and an highly reflective wrapping material, the Vikuiti TM Enhanced Specular Reflector by 3M~\footnote{\href{https://www.isoltronic.ch/assets/of-m-vikuiti-esr-app-guide.pdf}{Vikuiti/VM2000 data}} (formely known as VM2000), and filled with ZnWO$_4$ and BGO grains. The arrival time of the signals when the container was empty and filled with grains were also compared.

It has been observed that for the empty container, the difference in the charge collected when using an absorber or VM2000 is almost a factor of 3, while this difference is much less when the container is filled with grains \-- either BGO or ZnWO$_4$. This indicates that when the grains are present in the reference volume, a significant fraction of the light is captured by the WLS fiber before it reaches the VM2000.
The fraction of captured light when the container is filled with ZnWO$_4$ with respect to the amount of light captured with an empty container, is found to be 84 \%, indicating that a small fraction of the light is captured by the grains. For the same LED intensity the signal yield is smaller for ZnWO$_4$ grains than for BGO by about 15\%, but it is not expected to be a show-stopper for the performance required at FCC-ee.
Finally, with this small prototype, a preliminary study of the arrival time of the signal was performed comparing the configurations with the container (wrapped in black fabric) either empty or filled with grains. The average path length, $\Delta L$, is computed using the time difference, $\Delta T$, in the two configurations (empty or filled with scintillator grains) and the average refractive index of the medium, $n$: $\Delta L = c/n \times \Delta T$. Despite the two fibers being 4 mm away from each other, the average light path is expected to be large due to the multiple random reflections or refractions on each grain. Indeed, it is found to be of the order of 10 to 20 cm with ZnWO$_4$ grains.

\section{Test with cosmic muons}
\subsection{The 16-channels prototype \label{sec:protoDesc}}

\begin{figure}[htbp]
\centering
\includegraphics[width=.45\textwidth, trim = 0 -175 0 0 ]{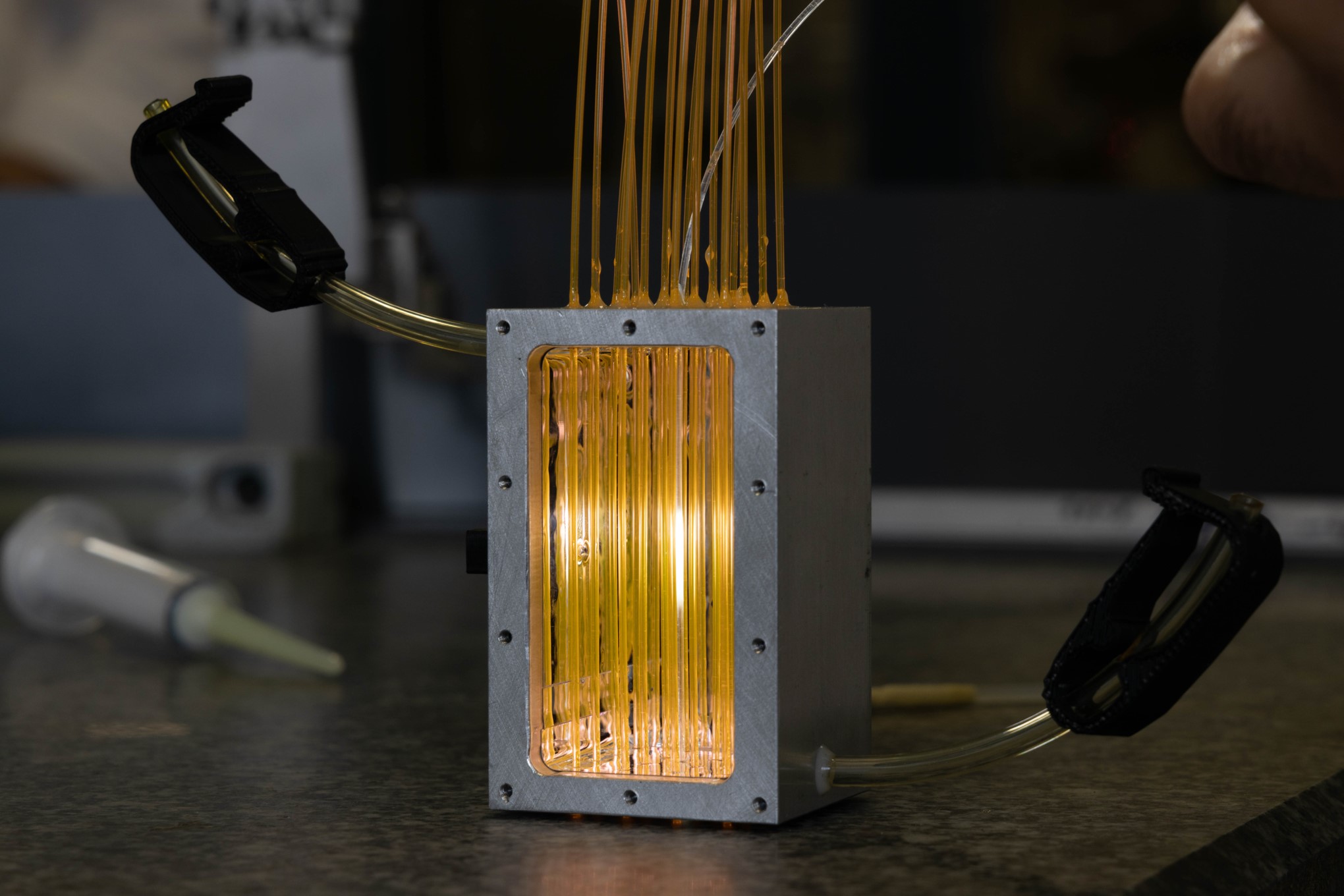}
\includegraphics[width=.42\textwidth]{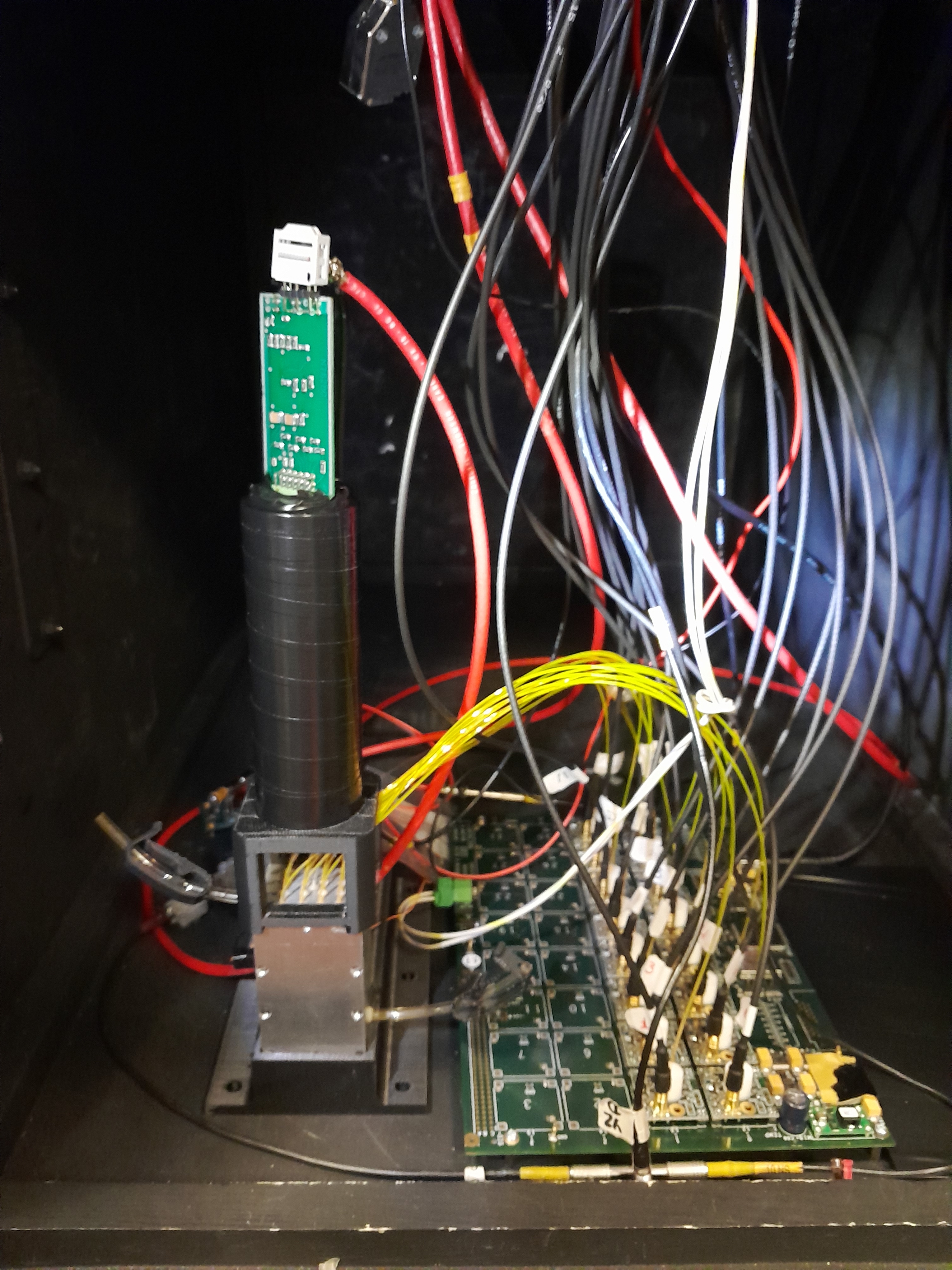}
\caption{The 16-channels GRAiNITA prototype (left) and the cosmic rays test bench (right). \label{fig:16channels}}
\end{figure}
The 16-channels GRAiNITA medium-size prototype is shown in figure~\ref{fig:16channels}-left. It consists of a metallic container with an active volume of $28 \times 28 \times 55$~mm$^3$ and it is equipped with 16 O-2(200) WLS fibers arranged in square geometry and inter-spaced by 7 mm. In order to monitor the behaviour of the setup, a clear fiber, unpolished along 1 cm, is placed in the middle of the WLS ones and connected to a green LED\footnote{Broadcom HLMP-CM1A-560DD, wavelength from 500 to 540 nm.} to inject the light in the active volume.
In order to study the detectors response homogeneity and to precisely measure the light yield per MeV, collected with ZnWO$_4$ and BGO grains, we developed a cosmic-ray test bench consisting of a small muon hodoscope. It is based on two 2 cm $ \times$ 2 cm plastic scintillators, read-out by Hamamatsu PMTs, placed immediately above and below the GRAiNITA detection volume, as shown in the right picture of figure~\ref{fig:16channels}. In the near future, we will add to the hodoscope, a TimePix detector, to study the light yield variation as function of the distance of the muon track to the closest fiber.

Each of the 16 WLS fibers of GRAiNITA is coupled to a SiPM mounted on a PCB for amplification and the signals from the SiPMs are digitized at 3.2 GigaSample/s with a WaveCatcher module~\cite{bib:WaveC} (see figure~\ref{fig:16channels}-right) for offline analysis. For BGO grains, the number of collected photoelectrons can be directly estimated via the integral of the SiPM signals over 2 $\mathrm {\mu s}$. On the other hand, for ZnWO$_4$ grains, due to the longer decay time constant of the scintillator, we implemented a dedicated version of both firmware and software of the WaveCatcher modules allowing to count the number of single photoelectron pulses over this long time scale ($\Delta t \sim  25 \ \mu \mathrm{s}$). The knowledge of the number of photo-electrons is of particular interest here, because the stochastic term of the relative energy resolution \-- beyond the fluctuations due to the energy deposition in the shower \--  depends on
the number of photo-electrons ($N_{\mathrm{PhE}}$): $\sigma_E/E \propto 1/\sqrt{N_{\mathrm{PhE}}}$.

\subsection{Prototype characterization}
Using the signal provided by the green LED connected to the clear fiber as described above, a first characterization of the prototype has been conducted. Events were recorded in three configurations: the prototype filled with ZnWO$_4$ grains only, with ZnWO$_4$ grains immersed in water and with ZnWO$_4$ grains immersed in ethylene-glycol. In future tests we plan to use a heavy liquid based on sodium tungstate dissolved in water for which the density is of the order of 3 (with a refractive index of about 1.85). The signals recorded have been corrected for the dark current background which is of the order of 2\%.
The density of the medium (ZnWO$_4$ + liquid or air)  in the prototype has been experimentally measured and found to be in good agreement with the expected values computed from the individual components.

\begin{figure}[htbp]
\centering
\includegraphics[width=.45\textwidth]{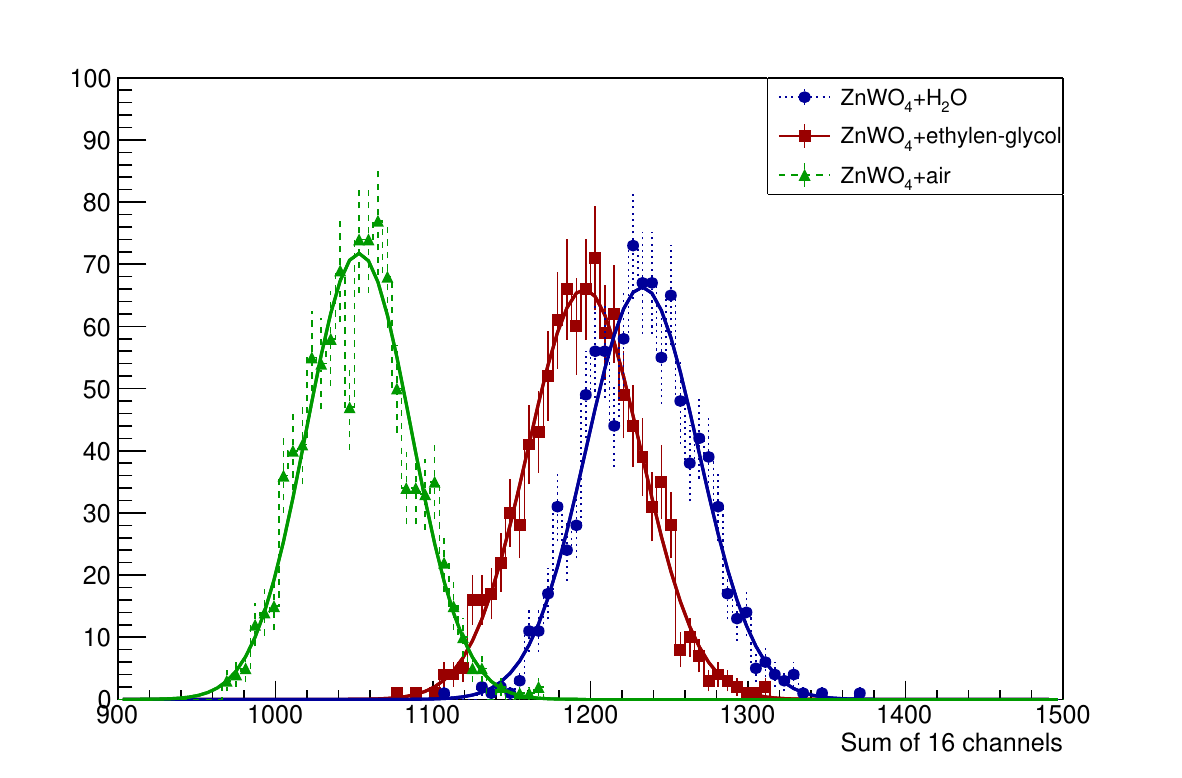}
\includegraphics[width=.45\textwidth]{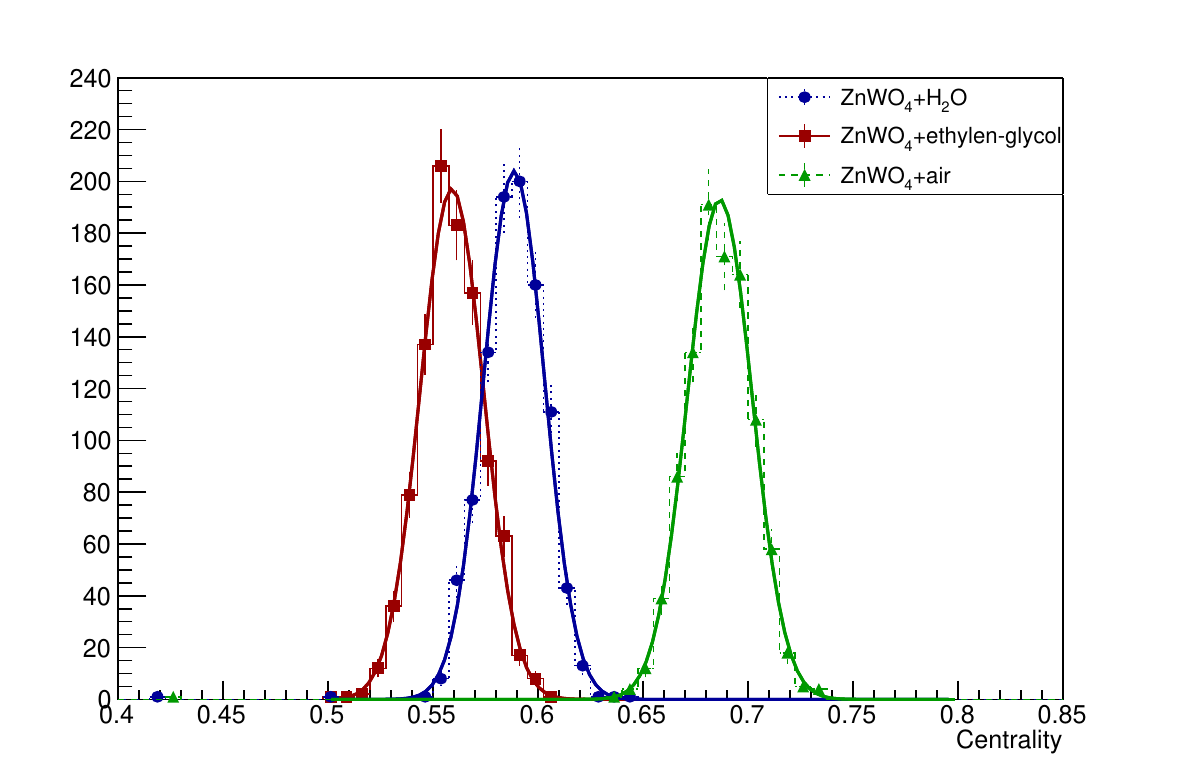}
\caption{Comparison of the sum of the signals recorded on the 16-channels (left plot) and the ratio of the signal recorded in the 4 more central fibers to the sum of the signal in the 16 fibers (right plot). The three data-taking configurations are shown: ZnWO$_4$ grains only (green dashed line), with ZnWO$_4$ grains immersed in water (blue dotted line) and with ZnWO$_4$ grains immersed in ethylene-glycol (red full line) with a green LED illumination (see text for details).\label{fig:GreenLED}}
\end{figure}

Two observables have been studied: the sum of the signals recorded on the 16-channels and the centrality of the event. The centrality is defined as the ratio of the signal recorded in the four more central fibers to the sum of the signal in the 16 fibers.
The results are shown in figure~\ref{fig:GreenLED} and summarized in table~\ref{tab:GreenLED} for the three configurations.

\begin{table}[htbp]
\smallskip
\begin{tabular}{l|l|l|l}
\hline
 External medium  & n(ZnWO$_4$)/n(Medium)  & Mean & Mean \\
& & (Sum) &  (Centrality) \\
  \hline
  Air (n=1) & 2.1  & $1052.9 \pm 1.1 $  & $0.6866 \pm 0.0005$ \\
Water (n=1.33)&  1.58 &  $1232.8 \pm 1.1 $& $0.5886 \pm 0.0005$  \\
Ethylene-glycol (n=1.43) & 1.47 & $1196.1 \pm 1.1$ & $0.5590 \pm 0.0005 $  \\
\hline
\end{tabular}
\caption{Measured mean values for the signal obtained from the sum of the 16 channels and for the centrality for the three configurations considered with the green light illumination (see text for details).\label{tab:GreenLED}}
\end{table}

The total amount of light collected results from the combination of three different effects: the amount of light which is emitted by the depolished part of the fiber (it was measured to be lower with liquids than in air), the smaller impact of the absorption in the grains when there is a better matching between the medium and the ZnWO$_4$ refractive indices, and the signal collection efficiency of the WLS fibers which varies with the medium refractive index.
The total amount of light collected by the system (Sum) with ZnWO$_4$ grains in air is about 12 to 15\% lower than with water or ethylene-glycol.

The slightly lower signal ($\sim 3 \%$) for the ethylen-glycol compared to water could be due to non-perfect transparency of the medium and to the presence of small bubbles which remain in the volume due to the higher viscosity of the liquid. It is of no-concern, including for the cosmic test, since, as it will be shown in sec.~\ref{sec:results}, the amount of photo-electrons is large enough even with ethylene-glycol.

The centrality of the signal represents how much the light is confined nearby its emission point. It is observed that the further apart the values of the ZnWO$_4$ and medium refractive indices, the more spatially confined is the signal. This can be attributed to the fact that with more different indices, the photons have higher probability to bounce back and to stay closer of the emission point, leading to a larger value of centrality.

\subsection{Results of the cosmic data-taking\label{sec:results}}
\begin{figure}[htbp]
\centering
\includegraphics[width=.45\textwidth]{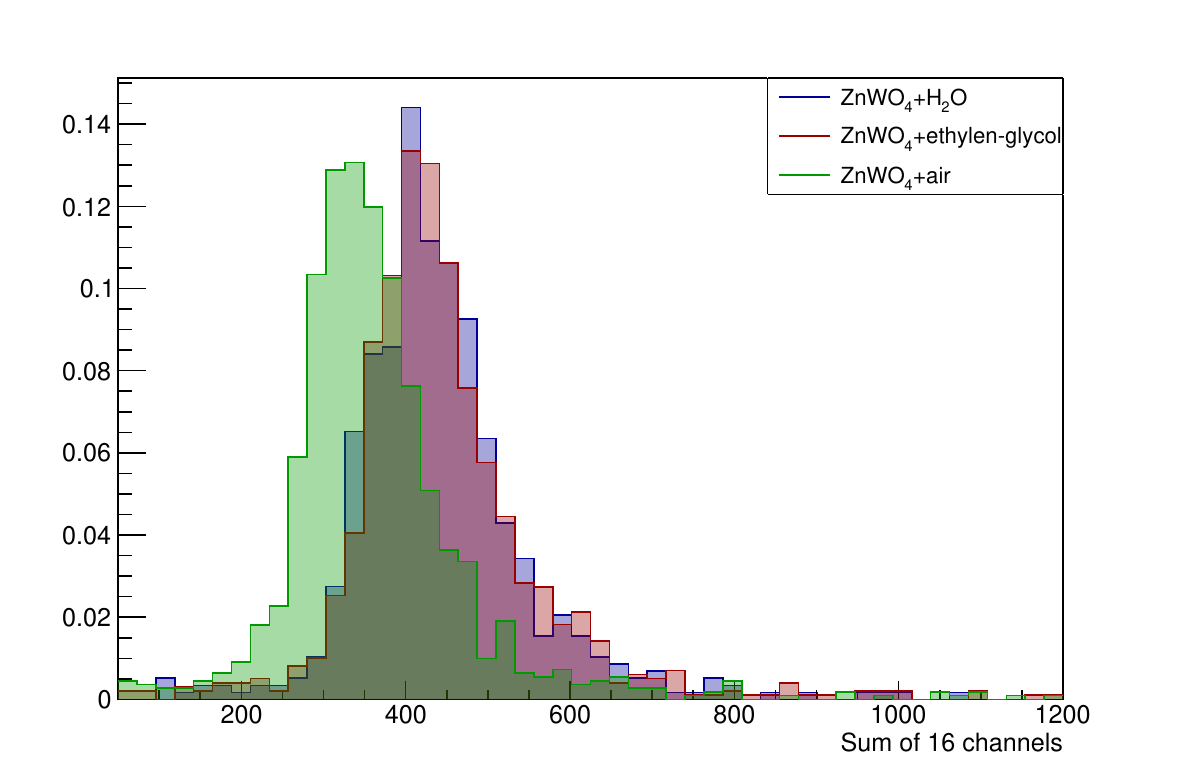}
\includegraphics[width=.45\textwidth]{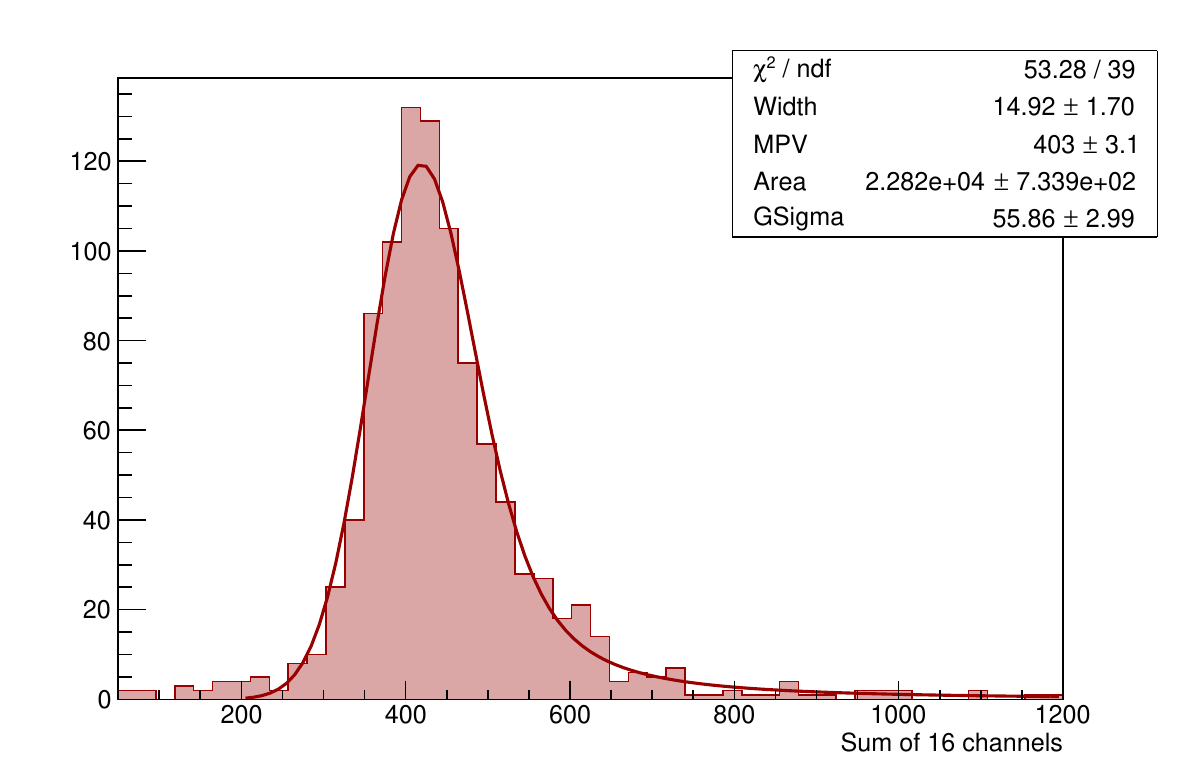}
\caption{Left: comparison of the sum of the signals recorded on the 16-channels (ZnWO$_4$ grains only, ZnWO$_4$ grains immersed in water and ZnWO$_4$ grains immersed in ethylene-glycol). Right: fit of the sum of the signals recorded on the 16-channels when the ZnWO$_4$ grains are immersed in ethylene-glycol (see text for details). \label{fig:Cosmic}}
\end{figure}

For this data-taking period, triggering is performed using the hodoscope described in section~\ref{sec:protoDesc}. Given the size of the prototype, the event rate is not very high: few events per hour. The same three configurations have been tested (Air, Water and Ethylene-glycol). The distributions of the sum of the 16 channels are shown in figure~\ref{fig:Cosmic}-left for the three cases ; about 500 events have been collected for each configuration. The signal is larger when the medium refractive index is better matched with the ZnWO$_4$ one. It is expected that the fraction of trapped light in the grain is larger with air, leading to a smaller signal. The sum of the 16 channels when the ZnWO$_4$ grains are immersed in ethylene-glycol has been fitted by a Landau function convoluted with a Gaussian function. From this fit, it is observed that about 400 photo-electrons are measured on average for a cosmic muon. Given the prototype characteristics,  the energy deposit of such muons amounts to about 40 MeV~\cite{bib:PDG,bib:NIST}. It is thus expected to collect of the order of 10000 photo-electrons per GeV with a GRAiNITA-like calorimeter. If this value is confirmed, it would open the road to a statistical fluctation of $1\%/\sqrt{E}$ on the energy resolution due to photo-electron statistics.

\section{Conclusion}

A very first prototype for a demonstrator of a new type of calorimeter technology using
 millimetric grains of ZnWO$_4$ immersed in a bath of transparent high-density liquid has been designed, constructed and tested using LED and cosmic rays. It was experimentally confirmed that, as expected, the signal remains confined in the vicinity of its production point and that a significant number of photoelectrons are created. These findings provide compelling evidence for a positive proof of concept for the GRAiNITA  design. In future, a full-size electromagnetic calorimeter module demonstrator will be built and characterized with test beam. Its dimensions of $17 \times 17 \times 40 \ \mathrm{cm}^3$, corresponding in depth to 25 X0, should allow to entirely contain a photon shower of up to 50 GeV energy.

\acknowledgments
We would like to thank Kuraray for providing fiber samples and useful discussions.


\bibliographystyle{JHEP}
\bibliography{main.bib}


\end{document}